\documentclass{aastex}
\usepackage{spr-astr-addons}
\usepackage{url}

\RequirePackage{color}

\begin{document}

\title{The phase relation between sunspot numbers and soft X-ray flares}
\shorttitle{Short article title}
\shortauthors{Yan et al.}

\author{X.L. Yan\altaffilmark{1,2}} \and \author{L.H. Deng\altaffilmark{1}}\and
\author{Z.Q ~Qu\altaffilmark{1}}
\and
\author{C.L Xu\altaffilmark{3}}

\altaffiltext{1}{National Astronomical Observatories/Yunnan
Astronomical Observatory, Chinese Academy of
              Sciences, Kunming, Yunnan 650011, P.R. China.}
\altaffiltext{2}{Graduate School of Chinese Academy of Sciences,
           Zhongguancun, Beijing, P.R. China.}
\altaffiltext{3}{Yunnan Normal University, Kunming, Yunnan,
P.R.China.}

\begin{abstract}
To better understand long-term flare activity, we present a statistical study on soft X-ray flares from May 1976 to May 2008. It is found that the smoothed monthly peak fluxes of C-class, M-class, and X-class flares have a very noticeable time lag of 13, 8, and 8 months in cycle 21 respectively with respect to the smoothed monthly sunspot numbers. There is no time lag between the sunspot numbers and M-class flares in cycle 22. However, there is a one-month time lag for C-class flares and a one-month time lead for X-class flares with regard to sunspot numbers in cycle 22. For cycle 23, the smoothed monthly peak fluxes of C-class, M-class, and X-class flares have a very noticeable time lag of one month, 5 months, and 21 months respectively with respect to sunspot numbers. If we take the three types of flares together, the smoothed monthly peak fluxes of soft X-ray flares have a time lag of 9 months in cycle 21, no time lag in cycle 22 and a characteristic time lag of 5 months in cycle 23 with respect to the smoothed monthly sunspot numbers. Furthermore, the correlation coefficients of the smoothed monthly peak fluxes of M-class and X-class flares and the smoothed monthly sunspot numbers are higher in cycle 22 than those in cycles 21 and 23. The correlation coefficients between the three kinds of soft X-ray flares in cycle 22 are higher than those in cycles 21 and 23. These findings may be instructive in predicting C-class, M-class, and X-class flares regarding sunspot numbers in the next cycle and the physical processes of energy storage and dissipation in the corona.
\end{abstract}



\section{Introduction}
Since the 11-year period of the solar cycle was discovered by Schwabe (1844), the long-term activity of the solar cycle has become a hot issue in the field of solar physics. Sunspots are taken as the most famous and typical indicators of solar activity. Sunspot activity has complex spatial and temporal behavior.
Carrington (1858, 1859) investigated a drift latitude of sunspot motion towards the equator and a variation of the rotation rate of the Sun.
Hathaway et al. (2003) examined the drift of the centroid of the sunspot area toward the equator in
each hemisphere from 1874 to 2002 and found that the drift rate slows as the centroid approaches the equator. The distribution of sunspots and flares in a solar cycle exhibits a ``butterfly diagram" (Carrington, 1858, Maunder 1904, 1913; Garcia 1990; Li et al. 2003) and the cycle appears uniformly
in both hemispheres on average (Newton \& Milson 1955; White \& Trotter 1977).

Sunspot activity is described by sunspot numbers and sunspot areas in general. Consequently, the sunspot numbers and sunspot areas are used to investigate long-term
solar activity (Li et al. 2009). Because of the periodicity of sunspot activity (sunspot numbers and sunspot areas), the associated eruptions (flares, CMEs) appear to have the similar periodicity to that of sunspot activity (Storini \& Hofer 1999). In addition, the soft X-ray flares were significantly delayed with respect to sunspot numbers, with a time lag of two to three years between the peak times in solar cycle 21 (Wagner 1988; Aschwanden 1994, Bromund et al. 1995). However, Wilson (1993) did not find evidence for a time lag between the maxima of the rates of optical flares and X-ray background flux in solar cycle 22. Wheatland \& Litvinenko (2001) found that there is an average delay of about 6 months considering cycles 21 and 22 together by using correlation analysis. For the cycle 23, Tan (2010) compared the sunspot numbers with the distributions of the appearance rate of solar GOES flares. They found that there is a time lag between the maximum value of sunspot numbers and the maximum values of the annual numbers of C-class, M-class, and X-class flares. The annual averaged relative sunspot number reaches its maximum in about 2000 and the annual numbers of C-class, M-class, and X-class flares reach their maxima in about 2001, 2001, and 2002 respectively. Temmer et al. (2003) found that there is a characteristic time lag between flare activity and sunspot activity in the range of 10-15 months for solar cycles 19, 21, and 23 by using the number of the flares in each month.

In order to investigate all kinds of long-term flare activity, the flare index calculated by T. Atac and A. Ozguc from Bogazici University Kandilli Observatory and the daily soft X-ray flare index defined by Antalov$\acute{a}$ (1996) are two main flare indices in the present research. For instance, Li et al. (2010) found that the northern-hemispheric flare activity should lead the southern-hemispheric flare activity for low-frequency components by using the former flare index. Joshi et al. (2004) reported a real north-south asymmetry during solar minimum and obtained the significant periods of approximately 28.26 days, 550.3 days, and 3.72 years by using the latter flare index.

In this paper, we focus on the three solar cycles (21, 22, and 23) to investigate the phase relation between sunspot numbers and flare activity (C-class, M-class, and X-class flares) by using a flare index, which is similar to Antalov$\acute{a}$ (1996).

\section{Observations and method}
The data used in this paper are as follows:

1. The monthly mean northern and southern hemispheric
sunspot numbers from January 1945 to December 2004 were compiled
by Temmer et al. (2006)($http://cdsweb.u-strasbg.fr/cgi-bin/qcat?J/A+A/447/735$). The monthly mean northern and southern hemispheric
sunspot numbers from January 1992 to December 2008 can be downloaded from NOAA's National Geophysical Data Center (NGDC) ($ftp://ftp.ngdc.noaa.gov/STP\protect\newline/SOLAR\_DATA/SUNSPOT\_NUMBERS\protect\newline/HEMISPHERE\_NUMBERS/$). There is an overlapping
time span from January 1992 through December 2004 with the
first sunspot data. The first group of data generally matches the second group (for details, see Temmer et al. 2006). In this paper, we extracted the sunspot numbers from May 1976 to May 2008 as our samples.

2. The soft X-ray flares were downloaded from NGDC ($ftp://ftp.ngdc.noaa.gov/STP/SOLAR\_DATA\protect\newline/SOLAR\_FLARES/FLARES\_XRAY/$).

In this paper, we adopted the peak fluxes of soft X-ray observed by GOES 1-8 \AA\ as the defined flare flux when one flare occurred.
Afterwards, we summed monthly peak fluxes of the defined flares (C-class, M-class, or X-class flares) as a flare index.
The equations are as follows:

\begin{equation}
F_C=\sum\limits_{i}n_i\times C_i\times10^{-3}
\end{equation}
\begin{equation}
F_M=\sum\limits_{i}n_i\times M_i\times10^{-2}
\end{equation}
\begin{equation}
F_X=\sum\limits_{i}n_i\times X_i\times10^{-1}
\end{equation}

In the above equations, $C_i$, $M_i$, $X_i$ indicate all sub-levels of C-class, M-class, X-class flares (e.g., For C-class flares, the sub-level flares ($C_i$) are C1.0, C1.1, C1.2, $\cdot\cdot\cdot\cdot\cdot\cdot$, and so on.). ``$i$" is given as a float number with one decimal (prior to 1980 April 27, ``$i$" is listed as an integer.). The value of ``$i$" is 1, 1.1, 1.2 $\cdot\cdot\cdot\cdot\cdot\cdot$, $j$. For C-class and M-class flares, the value of ``$i$" is no more than 9.9. $n_i$ denotes the monthly numbers of sub-level flares (e.g., $n_5$ denotes the monthly numbers of $C_5$, $M_5$ or $X_5$.). $F_C$, $F_M$, and $F_X$ indicate the monthly peak fluxes of C-class, M-class, and X-class flares. This method is the same as the flare index (FI) defined by Antalov$\acute{a}$ (1996) and used by Joshi et al. (2004). The unit of the peak flux is erg$\cdot$$cm^{-2}$$\cdot$$s^{-1}$.

As weak soft X-ray flares can not be detected during times of high activity (Feldman et al. 1997; Veronig et al. 2002), we discarded the classification of soft X-ray flares lower than C1.0 to remove the effect of the variation of soft X-ray background radiation.

\section{Results}
Figure 1 shows the distribution of the monthly numbers of three types of flares from May 1976 to May 2008. The period covers all of solar
cycle 21 (May 1976- July 1986), cycle 22 (August 1986- March 1996), and cycle 23 (April 1996- May 2008). The monthly flare numbers of C-class, M-class, and X-class flares have obvious periodical variability as well as sunspot numbers (see fig. 1). However, different flares have different peak intensities of soft X-ray flux. For instance, the peak flux of an X3.0 flare is 100 times larger than that of C3.0 flares. For M-class flares, the difference of the peak flux is about 10 times from M1.0 to M9.9. Considering the physics, we summed the monthly peak fluxes of all the sub-class flares of the same type flares. Detailed calculations can be seen from the equations in section 2. From table 1, the numbers of C-class flares are larger than that of M-class and X-class flares in each cycle. However, the total peak fluxes of the C-class flares are lower than those of M-class flares in cycles 21 and 22. Figs. 2a, b, c, d show that the monthly peak fluxes of the C-class, M-class, X-class flares and total peak fluxes of the three types of flares (blue lines) from May 1976 to May 2008 and their 13-point smoothed values superimposed (red lines). The monthly peak fluxes of C-class, M-class, and X-class flares have obvious periodical variability as well as the change of monthly sunspot numbers (Fig. 2e). To show the systematic time-lag (or time-lead) between the monthly peak fluxes of C-class, M-class, X-class flares and the monthly sunspot numbers in a cycle, we have done a correlation analysis of the two time series. In order to eliminate the impact that large, active flare-productive regions exert simultaneously to sunspot and flare numbers, we adopted the 13-point smoothed monthly peak fluxes of C-class, M-class, X-class flares and sunspot numbers to calculate the correlation coefficient and phase relation between them.

The smoothed monthly peak fluxes of C-class, M-class and X-class flares have positive correlation coefficients with the smoothed monthly sunspot numbers (seen from table 2). The correlation coefficients of M-class and X-class flares and sunspot numbers are higher in cycle 22 than those in cycles 21 and 23. Furthermore, the correlation coefficients of the smoothed peak fluxes between the three kinds of soft X-ray flares in cycle 22 are higher than those in \textbf{21} and 23. The correlation coefficients between the smoothed monthly sunspot numbers and the smoothed monthly peak fluxes of C-class, M-class, and X-class flares can be seen in figure 3. The abscissa indicates the shift of the smoothed monthly sunspot numbers pertaining to the smoothed monthly peak fluxes of C-class, M-class and X-class flares, with negative values representing backwards shifts. It is found that the monthly peak fluxes of C-class, M-class, and X-class flares have a very noticeable time lag of 13, 8, and 8 months respectively in cycle 21 with respect to sunspot numbers. There is no time lag between the sunspot numbers and M-class flares in cycles 22. However, there is a one-month time lag for C-class flare and a one-month time lead for X-class flare with respect to sunspot numbers in cycle 22. Moreover, the smoothed monthly peak fluxes of C-class, M-class, and X-class flares have a very noticeable time lag of one month, 5 months, and 21 months respectively with respect to the smoothed monthly sunspot numbers in cycle 23. If we take the three types of flares together, there is an obvious time lag of 9 months in cycle 21, no time lag in cycle 22 and a characteristic time lag of 5 months in cycle 23 with respect to sunspot numbers.

\section{Conclusion and Discussion}
In this paper, we present a statistical study on three types of soft X-ray flares from May 1976 to May 2008. We use the data of the smoothed monthly peak fluxes of C-class, M-class, and X-class flares and the smoothed monthly sunspot numbers. The main results are as follows:

1. The smoothed monthly peak fluxes of C-class, M-class, and X-class flares have a very noticeable time lag of 13, 8, and 8 months respectively with respect to the smoothed monthly sunspot numbers in cycle 21.

2. There is a one-month time lag for the smoothed monthly peak fluxes of C-class flares, a one-month time lead for the smoothed monthly peak fluxes of X-class flares and no time lag for the smoothed monthly peak fluxes of M-class flares with respect to the smoothed monthly sunspot numbers in cycle 22.

3. The smoothed monthly peak fluxes of C-class, M-class, and X-class flares have a very noticeable time lag of one month, 5 months, and 21 months in cycle 23 respectively with respect to the smoothed monthly sunspot numbers.

4. If we take the three types of flares together, we find that the soft X-ray flares have an obvious time lag of 9 months in cycle 21, no time lag in cycle 22 and a characteristic time lag of 5 months in cycle 23 with respect to the smoothed monthly sunspot numbers.

5. The correlation coefficients of the smoothed monthly peak fluxes of M-class and X-class flares and the smoothed monthly sunspot numbers are higher in cycle 22 than those in cycles 21 and 23.

6. The correlation coefficients between the three kinds of soft X-ray flares in cycle 22 are higher than those in \textbf{21} and 23.

In this paper, we adopted the method developed by Antalov$\acute{a}$ (1996) and used the smoothed monthly peak fluxes of soft X-ray flares. We divided the soft X-ray flares into three groups according to the classification of soft X-ray flare, which is slightly different from Temmer et al. (2003). When the numbers of flares are multiplied by the peak fluxes, there is a great change in value pertaining to the original flare numbers. From table 1, one can see that the numbers of M-class, X-class flares are small but they have large monthly peak fluxes. The peak fluxes of flares can be approximately taken as the energy released by flares. As M-class and X-class flares have released more energy than that of B-class and C-class flares, we think that the peak fluxes of flares taken as the flare index may be closer to the energy released by the flares. That is why we used this method to analyze the phase relation.

The activity of the C-class, M-class and X-class flares exhibited an obvious time lag behind sunspot activity in odd-numbered cycles, while in even-numbered cycles M-class flares do not. Also, the C-class and X-class flares show reversed behavior in the even cycle. If we consider the three types of flares together, we find that there is an obvious time lag in the odd solar cycles and no time lag in the even solar cycle. Though different sorts of flares show different time lags or time leads with respect to sunspot numbers, there is a regular time delay in the odd solar cycles and no time delay in the even solar cycle between soft X-ray flares and sunspot numbers after we sum the three types of flares together. The value of summing the monthly peak fluxes of three types of flares approximately denotes the total energy released by flares. These results imply that the flare activities are related to the 22-year magnetic cycle of the Sun. We need forthcoming data to confirm these results and model the process of flare energy storage and dissipative mechanism.

\begin{table*}{}
\begin{center}
TABLE 1 \\
The numbers and peak fluxes of three types of soft X-ray flares (C-class, M-class, and X-class flares) in cycles 21, 22, and 23. The values in parentheses denote the peak fluxes of the defined flares. ``Total" denotes the total numbers and the total peak fluxes in the selected period from May 1976 to May 2008. The unit of flare peak flux is erg$\cdot$$cm^{-2}$$\cdot$$s^{-1}$.\\
\vspace{0.5cm}
\begin{tabular}{llllllll}
\tableline\tableline \label{tbl-1}
Flare class &21 cycle  &22 cycle & 23 cycle  &Total     \\
\tableline
C &14576 (46.76)&12430(41.79) &13148 (39.39)&40154(127.94)\\
M& 2175(50.22)&2019(48.15)&1439(34.37)&5633(132.74)\\
X&165(37.59)&152(35.46)& 126(31.08)&443(104.13) \\
\tableline
\end{tabular}
\end{center}
\end{table*}

\begin{table*}{}
\begin{center}
TABLE 2 \\
Correlation coefficients between the smoothed monthly peak fluxes of C-class flares, M-class flares, X-class flares and the smoothed monthly sunspot numbers in cycle 21 (the values in parentheses denote correlation coefficients in cycles 22 and 23.). ``SSN", ``SFC"  ,``SFM", and ``SFX" indicate the smoothed monthly sunspot numbers, the smoothed monthly peak fluxes of C-class, M-class, and X-class flares respectively.
 \\
\vspace{0.5cm}
\begin{tabular}{lllllll}
\tableline\tableline \label{tbl-1}
& SFC  & SFM  & SFX  \\
\tableline
SSN &0.751(0.974, 0.984)  &0.825(0.950, 0.908) & 0.778(0.921, 0.457)           \\
SFC &&  0.951(0.954, 0.948)&0.874(0.908, 0.478) \\
SFM &&& 0.959(0.974, 0.655)\\

\tableline
\end{tabular}
\end{center}
\end{table*}

\begin{figure*}
   \centering
   \includegraphics[width=12cm]{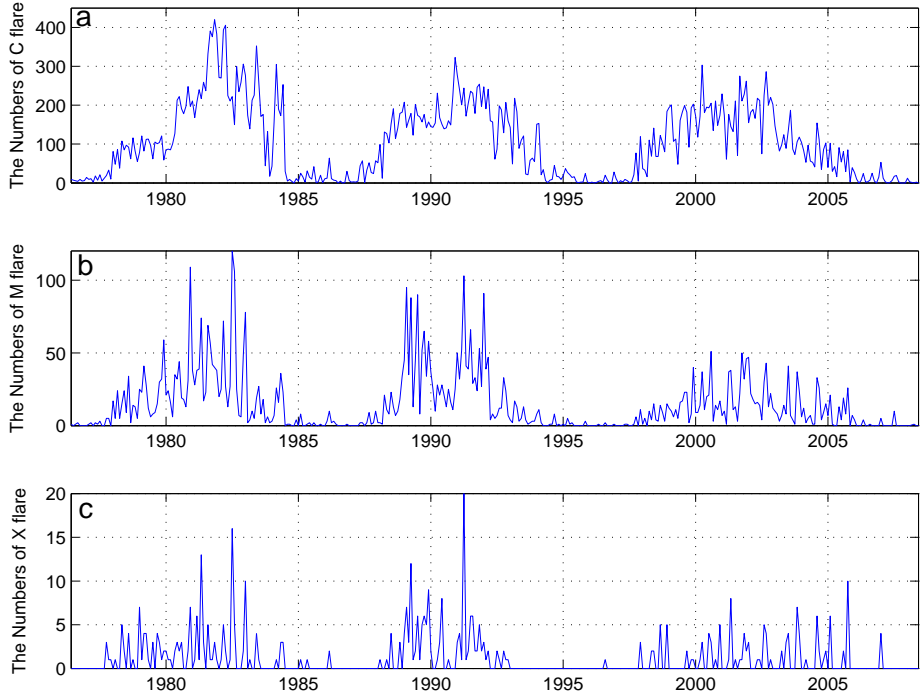}
\caption{The distribution of the monthly numbers of three types of flares from May 1976 to May 2008 respectively.}
         \label{}
   \end{figure*}

\begin{figure*}
\centering
   \includegraphics[width=12cm]{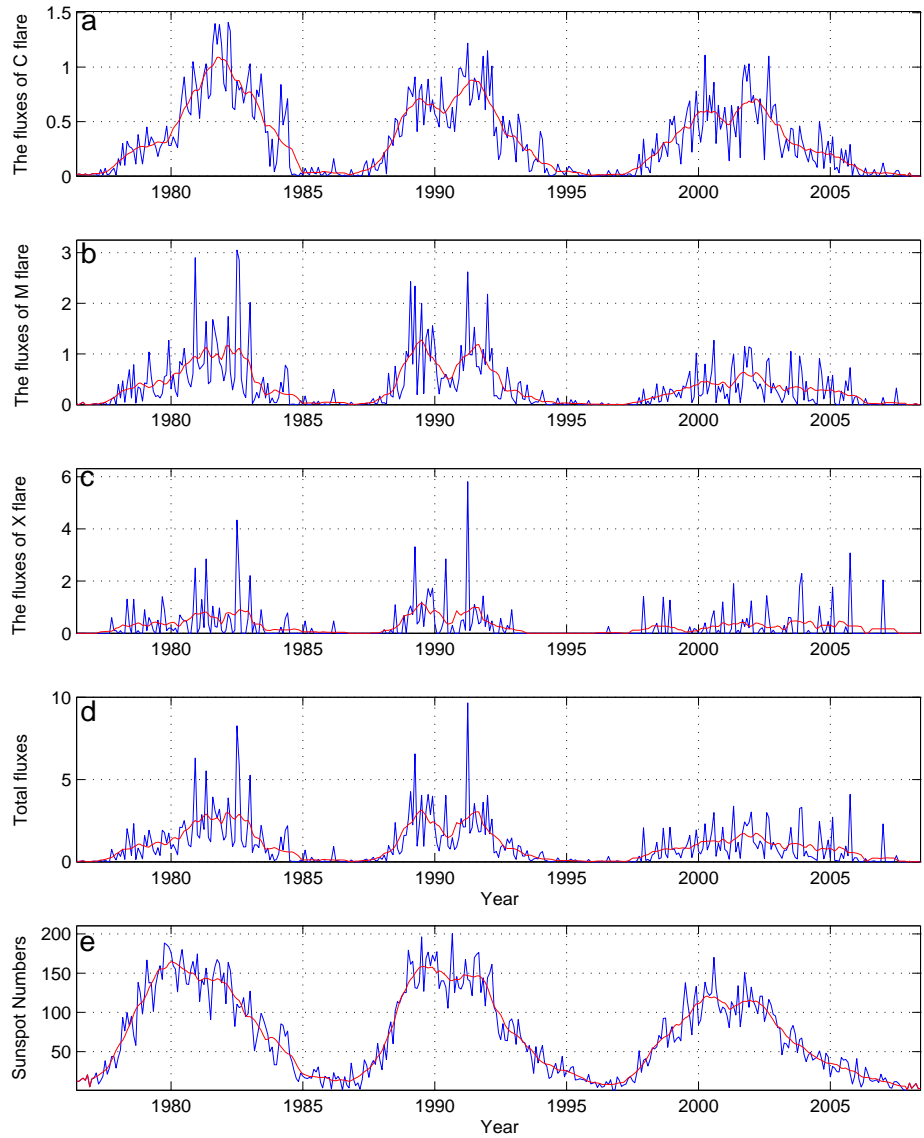}
\caption{The distribution of the monthly peak fluxes of C-class (a), M-class (b), X-class (c) flares, total peak fluxes of three types of flares(d), the monthly sunspot numbers(e) (blue lines) and their 13-point smoothed values (red lines) superimposed.}
\end{figure*}

\begin{figure*}
\centering
   \includegraphics[width=12cm]{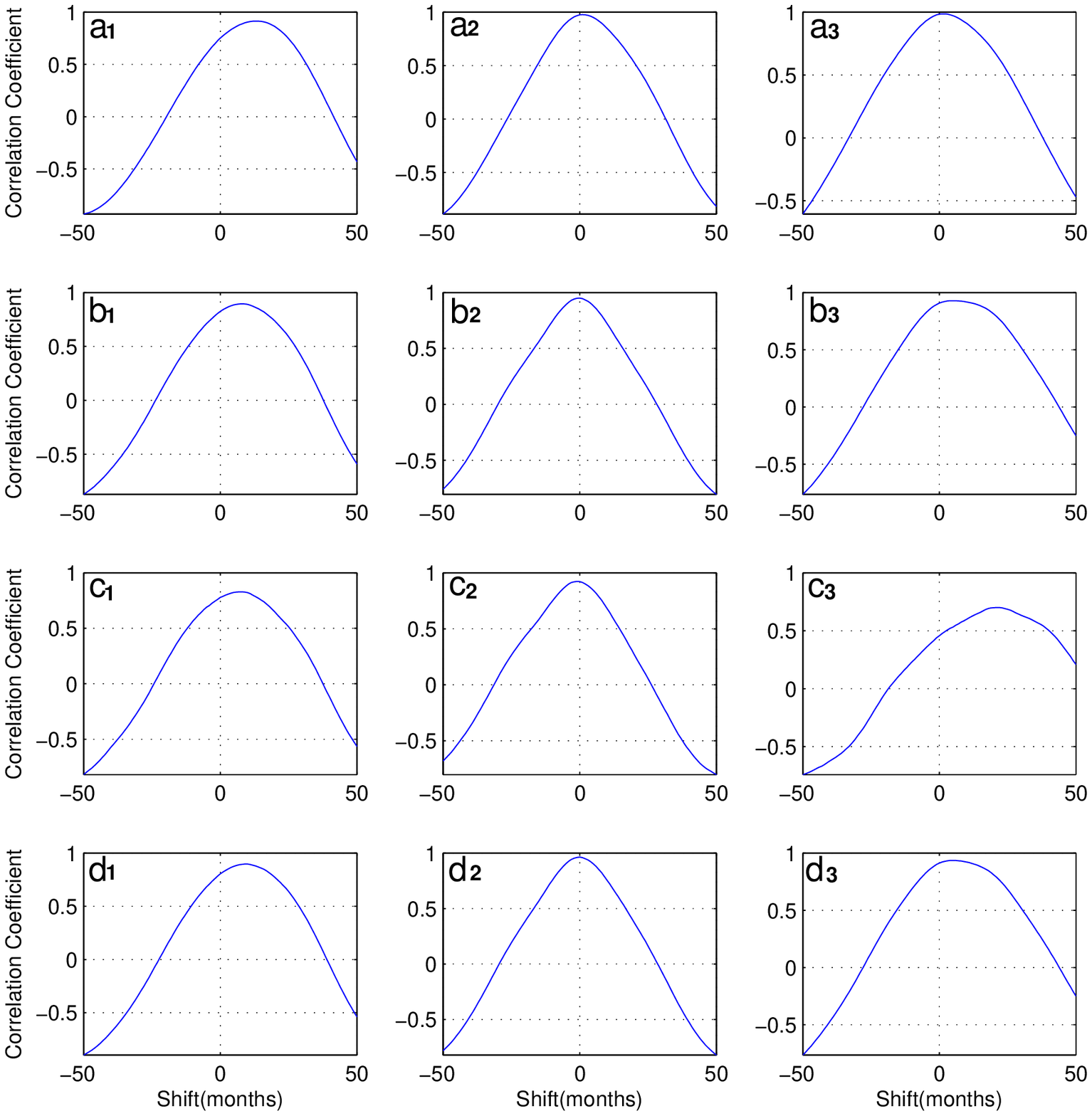}
\caption{Cross-correlation coefficients between the smoothed monthly peak fluxes of C-class ($a_1$-$a_3$), M-class ($b_1$-$b_3$), X-class ($c_1$-$c_3$) flares, the total peak fluxes of the three types of flares ($d_1$-$d_3$) and the smoothed monthly sunspot numbers in cycles 21, 22, and 23.
The abscissa indicates the shift of the smoothed monthly sunspot numbers with respect to the smoothed monthly peak fluxes of C-class, M-class, and X-class flares, with negative values representing backwards shifts.}
\end{figure*}

\acknowledgments
The authors thank the anonymous referee for her/his careful reading of the manuscript and constructive
comments that improved the original version. The authors thank the staff of all the Web sites that provide the data for the public to download. This work is supported by the National Science Foundation of
China (NSFC) under grant numbers 10903027, 10673031 and 40636031, 11003041, Yunnan Science Foundation of China under grant number 2009CD120, and the
National Basic Research Program of China 973 under grant number.


\end{document}